\documentclass[journal]{IEEEtran}
\usepackage{amsmath,amssymb,graphicx,multirow,array,enumitem,adjustbox}
\usepackage{booktabs, threeparttable,caption, makecell}
\usepackage{stfloats}
\usepackage{makecell}
\usepackage{pifont,wasysym}

\usepackage{bbm}
\usepackage[ruled, lined, linesnumbered, commentsnumbered, longend]{algorithm2e}
\usepackage[colorlinks=true,citecolor=green]{hyperref}
\usepackage{cleveref}
\usepackage{color}

\ifCLASSINFOpdf
\else
\fi

\begin{document}
%
\title{Self-Supervised Learning with Cluster-Aware-DINO for High-Performance Robust Speaker Verification}
%
%
%

\author{Bing Han, \IEEEmembership{Student Member, IEEE,}
        Zhengyang Chen, \IEEEmembership{Student Member, IEEE,}\\
        and Yanmin Qian, \IEEEmembership{Senior Member, IEEE}
\thanks{
Part of the results have been presented at Interspeech 2022 \cite{han2022self}.

All the authors are with the X-Lance Lab, Department of Computer Science and Engineering \& MoE Key Laboratory of Artificial Intelligence, AI Institute, Shanghai Jiao Tong University, Shanghai, 200240 P. R. China (e-mail:\{hanbing97, zhengyang.chen, yanminqian\}@sjtu.edu.cn)}

}

%
%

\markboth{Journal of \LaTeX\ Class Files,~Vol.~14, No.~8, August~2015}%
{Shell \MakeLowercase{\textit{et al.}}: Bare Demo of IEEEtran.cls for IEEE Journals}
%



\maketitle

\begin{abstract}
Automatic speaker verification task has made great achievements using deep learning approaches with the large-scale manually annotated dataset. 
However, it's very difficult and expensive to collect a large amount of well-labeled data for system building. 
Recently, self-supervised speaker verification has attracted a lot of interest by the reason of its no-dependency on labeled data. 
In this article, we propose a novel and advanced self-supervised learning framework which can construct a very strong speaker verification system with high performance without using any labeled data. 
To avoid the impact of false negative pairs from the contrastive-learning based self-supervised learning, we adopt the self-distillation with no labels (DINO) framework as the initial model, which can be trained without exploiting negative pairs. 
Then, we further introduce a cluster-aware training strategy for DINO to improve the diversity of data.
In the iteration learning stage, due to a mass of unreliable labels from unsupervised clustering, the quality of pseudo labels is important for the system performance. This motivates us to propose dynamic loss-gate and label correction (DLG-LC) methods to alleviate the performance degradation caused by unreliable labels. More specifically, we model the loss distribution with Gaussian Mixture Model (GMM) and obtain the loss-gate threshold dynamically to distinguish the reliable and unreliable labels.
Besides, we adopt the model predictions to correct the unreliable label, for better utilizing the unreliable data rather than dropping them directly.
Moreover, we extend the DLG-LC from single-modality to multi-modality on the audio-visual dataset to further improve the performance.
The experiments are performed on the commonly used Voxceleb dataset. Compared to the best-known self-supervised speaker verification system, our proposed method obtain 22.17\%, 27.94\% and 25.56\% relative EER improvement on Vox-O, Vox-E and Vox-H test sets, even with fewer iterations, smaller models, and simpler clustering methods.
More importantly, the newly proposed self-supervised learning system even achieves comparable results with the fully supervised system on Voxceleb dataset, but without using any human labeled data.

\end{abstract}


\IEEEpeerreviewmaketitle

\section{Introduction}

\IEEEPARstart{S}{eaker} verification (SV) is a task that utilizes speech as the biometric feature to verify the speakers’ identities. 
Recently, deep learning methods have been widely applied for speaker verification (SV) tasks and many efforts have been made such as various model architecture~\cite{dvector, xvector, rvector, han2022local, han2022mlp}, training objection~\cite{aam, chung2020defence, heigold2016end}, pooling methods~\cite{revising_pooling, zhu2018self} and so on, to achieve excellent performance compared with traditional methods such as Gaussian Mixture Model-Universal Background Model (GMM-UBM)~\cite{gmmubm}, i-vector~\cite{ivector}.
However, all these methods are based on fully-supervised training and usually require large amounts of training data with accurate human annotations, while as we know that the collection of large-scale well-labeled data is actually very difficult and expensive. 

To reduce the high dependency on labeled data, recently self-supervised learning has attracted a lot of interest and some researchers are focusing on applying it to speaker verification tasks. 
Inspired by the great success of speech pre-trained models, e.g. wav2vec 2.0~\cite{baevski2020wav2vec} and HuBERT~\cite{hsu2021hubert}, in automatic speech recognition (ASR) tasks, some researchers~\cite{fan2020exploring} tried to extract the universal speech representation to fine-tune on SV task directly. 
Since these pre-trained models are trained without explicit speaker information, the results of simply fine-tuning are not ideal. 
In the work~\cite{chen2022large}, the speech representation learned from large-scale unlabeled data were explored to replace the acoustic features, and then the normal supervised deep model was trained as usual. Although promising performance has been obtained, it still requires labeled data for training and the parameter size is unacceptable for real applications due to the large pre-trained model.

To take full advantage of the large-scale unlabeled data, inspired by text-to-speech (TTS) task, a generative method has been investigated in~\cite{stafylakis2019self} to separate speaker representation with the help of phone information. 
Subsequently, some researchers came up with a hypothesis that speech segments truncated from the same utterance belong to the same speaker while those from different utterances belong to different speakers, which is approximately true. 
Based on this hypothesis, many efforts~\cite{inoue2020semi, huh2020augmentation, zhang2021contrastive, xia2021self, mun2020unsupervised} have been made to obtain discriminative speaker representations by maximizing information between different segments from the same utterance via contrastive-learning. 
Then, inspired by~\cite{caron2018deep}, an iterative learning framework~\cite{cai2021iterative} was developed to further improve the performance of self-supervised SV system. 
This state-of-the-art system usually consists of two stages. In the first stage, contrastive-learning based objective function is applied to train a speaker encoder. In stage II, it adopts the pre-trained model in stage I to estimate the pseudo labels by clustering and then uses them as the supervised signal to train a new encoder. This process is performed iteratively to improve the performance continuously. 

This two-stage framework has obtained excellent performance~\cite{idlab2020,jhu2021,SNU2021,tao2021self,cai2022incorporating}, but there are many shortcomings, which restrict the further improvement of the system performance. 
For contrastive-learning methods in stage I, speech segments cropped from different utterances are regarded as negative pairs to be pushed away from each other in speaker space.
However, different utterances may belong to the same speaker in the real situation, which shows that this inaccurate assumption might make some mistakes. 
For the second iterative stage, \cite{caron2018deep,cai2021iterative} have proved that many pseudo labels generated by the clustering algorithm lack reliability, which would confuse and degrade the model. 
Hence, the key to improving the performance is finding a way to select high-quality pseudo labels.
Based on this hypothesis, in~\cite{tao2021self}, they observed that the data with lower loss is more reliable than those with unreliable labels, and then proposed a loss-gate learning strategy to distinguish the reliable labels from unreliable ones by setting a loss threshold. Only the data whose loss is under the threshold can be used to update the network.
Although this approach led to further improvements, the manually set thresholds in each iteration are not flexible, and data with unreliable labels is not fully utilized.

In this paper, we propose several new strategies for self-supervised learning speaker verification. Firstly, we introduce the DINO (distillation with no labels)~\cite{dino} in the first pre-training stage, which is only based on maximizing the similarity between the augmented segments pairs sampled from the same utterance. To minimize channel and environmental impacts and increase data's diversity, we propose a cluster-aware training strategy for DINO to further improve its performance. 
In the second iterative stage, we model the loss distribution data using GMM with two components, in which each component represents the data with reliable labels or unreliable labels. Then, the dynamic loss-gate (DLG) threshold, computed with the estimated GMM, is used to distinguish the two types of labels, which is more flexible than the manually tuned threshold. Besides, inspired by semi-supervised learning works~\cite{berthelot2019remixmatch,sohn2020fixmatch,zhang2021flexmatch}, we propose label correction (LC) to leverage the model's prediction as target label and use it to correct the unreliable pseudo label, instead of discarding the unreliable data directly~\cite{tao2021self}. 
Finally, we incorporate multi-modality into the above proposed DLG-LC strategy and clustering step. Benefiting from the complementary audio and visual information of different modalities, DLG-LC can select the data with reliable labels more effectively. 



The main contributions of this paper are summarized as follows:
\begin{enumerate}
    \item The DINO framework is introduced as the self-supervised learning framework to obtain the initial pre-trained model, which is negative-pairs free to avoid the impact of the false negative pairs. In addition, cluster-aware training strategy is designed to enhance DINO and it can improve the diversity of data and obtain better performance. 
    \item To select the high-quality data more effectively and flexibly in the second iterative stage, dynamic loss-gate (DLG) is developed which can determine the loss-gate threshold dynamically to select the data with reliable labels. Meanwhile, label correction (LC) is also adopted to further improve the results.
    \item The DLG-LC method is further extended from audio single-modality to audio-visual multi-modality. Multi-modal data utilize multi-modal knowledge and make reliable label selection more efficient.
    \item With these strategies, we achieve a great performance leap compared with the state-of-the-art (SOTA) system with self-supervised learning nowadays, even with fewer iterations, smaller models, and simpler clustering methods. More promisingly, this newly proposed self-supervised learning framework can approach the current SOTA of the fully supervised system, and achieve comparable performance.
\end{enumerate}

\section{Self-Supervised Learning for Speaker Verification}
In this section, the commonly utilized two-stage self-supervised speaker verification framework is reviewed, including the first contrastive-learning stage for pre-trained model and the second iterative learning stage. 

\subsection{Contrastive based Self-Supervised Speaker Verification}
Self-supervised learning (SSL) is a type of unsupervised training manners, which can design pretext or proxy and learn the representations from the data itself. Common SSL methods can be roughly divided into two classes: generative~\cite{stafylakis2019self} and contrastive~\cite{nagrani2020disentangled,chung2020seeing} methods. 
Based on the hypothesis that segments sampled from the same utterance belong to the same speaker while those from different utterances come from different speakers, most studies of SV tasks focus on contrastive learning approaches. 
Among them, SimCLR~\cite{chen2020simple} is one of the most popular contrastive learning frameworks. Its basic idea is that minimize the distance between the representations of augmented segments cropped from the same utterance as well as maximize negative pairs from different utterances. 
Besides, MoCo~\cite{MOCO} framework provides further performance gain through a dynamic dictionary with a queue and a moving-averaged encoder. 
Based on these frameworks, many works such as equilibrium learning~\cite{mun2020unsupervised}, augmentation adversarial training~\cite{huh2020augmentation}, channel-invariant training~\cite{zhang2021contrastive}, prototype momentum~\cite{xia2021self} are proposed to learn more discriminative speaker representation. 

\subsection{Iterative Framework for Self-Supervised Speaker Verification}
Considering that the assumption of contrastive learning can naturally introduce label error and might degrade the model, 
in~\cite{cai2021iterative}, they
proposed an iterative, self-evolving framework to further improve the performance of self-supervised speaker verification systems. This framework is mainly divided into two stages, and they are illustrated as follows:
\begin{itemize}
    \item Stage I: Pre-training
    \begin{enumerate}
        \item Use contrastive learning or other self-supervised learning methods to pre-train a speaker encoder as the initial model.
        \item With the pre-trained model, extract the speaker embeddings for the training set and then apply a clustering algorithm to assign pseudo labels.
    \end{enumerate}
    \item Stage II: Iterative training and pseudo labeling.
    \begin{enumerate}
        \item Train a new encoder with the pseudo labels generated by the previous step.
        \item Perform a clustering algorithm to update pseudo labels with the new encoder.
        \item Repeat stage II several times until the model converges.
    \end{enumerate}
\end{itemize}

Although this framework requires high computing resources due to the several iterations, it is widely used in ~\cite{idlab2020,jhu2021,dku2021,SNU2021,tao2021self} for its advanced performance. In addition, this framework is extended to the audio-visual dataset in~\cite{cai2022incorporating} and achieves better performance with the help of multi-modal information in the clustering algorithm.

\section{Cluster-Aware-DINO for Speaker Verification}

\label{sec:audio_visual_net}

\begin{figure*}[ht]
  \centering
  \includegraphics[width=0.9\linewidth]{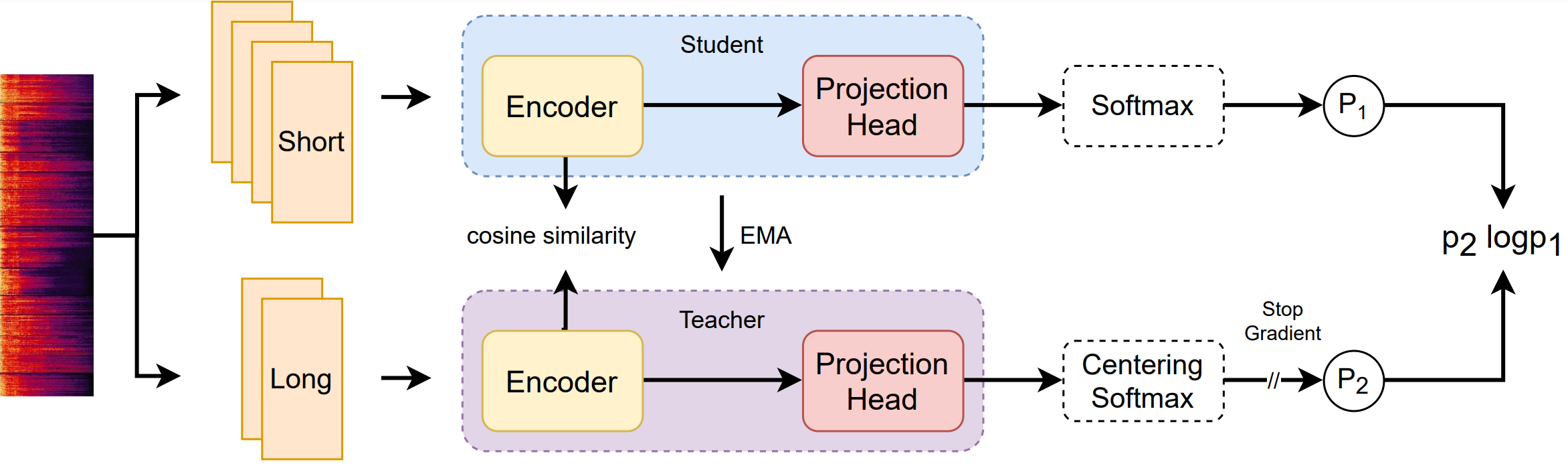}
  \caption{Framework of distillation with no label (DINO) for self-supervised speaker representation learning}
  \label{fig:dino}
\end{figure*}

For contrastive-learning based methods in previous work, they shared the same assumption that segments in a batch belong to different speakers. 
But this assumption does not hold all the time because repeat speakers might appear in the same batch. Taking the statistics on Voxceleb 2 as an example, we can compute the probability of repeat speakers on Voxceleb 2 by Equation.\ref{equ:prob} and the results are listed in Table.~\ref{tab:negative_pair_prob}. 
\begin{equation}
\label{equ:prob}
    p_{repeat}(S, N) = 1-\frac{A_S^N}{S^N} = 1-\frac{S!}{S^N(S-N)!} 
\end{equation}
where $S$ is the speaker number in training set and $N$ is batch size. 
\begin{table}[ht]
  \caption{The probability of repeat speaker in a batch}
  \label{tab:negative_pair_prob}
  \centering
  \begin{tabular}{c|ccccc}
    \toprule
    \textbf{Batch Size} & 16 & 32 & 64 & 128 & 256 \\ \midrule
    \textbf{Probability} & 0.020 & 0.080 & 0.286 & 0.745 & 0.996 \\
    \bottomrule
  \end{tabular}
\end{table}

According to the table, 
the larger batch size leads to a higher probability of repeating which will cause a bad impact on the model. We can use a small batch size to alleviate this problem, but it will degrade the performance~\cite{MOCO}.

\subsection{DINO based Self-Supervised Learning}

To tackle this problem, negative-pairs free DINO~\cite{dino} is introduced to self-supervised speaker verification task, and the whole framework is shown in Fig.~\ref{fig:dino}. 

Firstly, 4 short $\{x_1^s, x_2^s, x_3^s, x_4^s\}$ and 2 long segments $\{x_1^l,x_2^l\}$ are randomly sampled from an utterance using a multi-crop strategy~\cite{caron2020unsupervised}, and the long segments can be used to extract more stable speaker embedding. It is notable that when sampling, these segments should overlap as little as possible.
Same as the previous work~\cite{huh2020augmentation, zhang2021contrastive, xia2021self, mun2020unsupervised}, we still obey the assumption that the segments cropped from the same utterance belong to the same speaker and then apply different kinds of data augmentation on them by adding noise or room impulse response for robust performance. 
Unlike SimCLR~\cite{chen2020simple}, which only uses one encoder to do contrastive learning, our model consists of not only a \textit{student} encoder but also a momentum \textit{teacher} encoder whose architecture is similar as knowledge distillation~\cite{hinton2015distilling}. After augmentation, all segments pass through the \textit{student} while only the long segments pass through the \textit{teacher}, thus encouraging the \textit{short-to-long} correspondences by minimizing the cross-entropy $H(\cdot)$ between two distributions as the following Equation.\ref{equ:ce}:
\begin{equation}
\label{equ:ce}
L_{ce} = \sum_{x\in \{x_1^l,x_2^l\}}^{} \sum_{x'\in \{x_1^l,x_2^l,x_1^s, \dots, x_4^s\}}^{} H(P_t(x) \mid P_s(x')) 
\end{equation}
where output distributions of momentum \textit{teacher} network $f_{\theta_t}$ and \textit{student} network $f_{\theta_s}$ are denoted by $P_t$ and $P_s$ respectively. And $P$ can be computed by using a softmax function to normalize the output:
\begin{equation}
\label{equ:sharpen}
P_s(x) = Softmax(\frac{f_{\theta_s}(x)}{\epsilon_s} )
\end{equation}
where $\epsilon_s>0$ is the temperature parameter that can control the sharpness of the output distribution. Similarly, there is a formula holds for $P_t$ with temperature $\epsilon_t > 0$, too. Moreover, a mean computed over batches is used for centering \textit{teacher} model's output distribution. During the training, both sharpening and centering are applied to avoid trivial solution~\cite{dino}.

The \textit{teacher} and \textit{student} own the same architecture but with different parameters due to the different update methods. The \textit{student} is updated by gradient descent while the \textit{teacher} is updated by the exponential moving average (EMA) of the \textit{student}'s parameters. EMA's update rule is:
\begin{equation}
    \label{equ:ema}
    \theta_t \leftarrow \lambda \theta_t + (1-\lambda) \theta_s
\end{equation}
where $\lambda$ is adjusted by a cosine schedule~\cite{loshchilov2016sgdr} from $0.996$ to $1$ during training. 
Speaker embeddings are extracted by Encoders. Then speaker embeddings are fed into the Projection Head which contains a 3-layers perceptron with hidden dimension 2048 followed by $\ell_2$ normalization and a weight normalized fully connected layer with \textit{K} dimensions. The whole architecture is similar to~\cite{caron2020unsupervised}. 

In addition, a cosine-based consistency loss is added to ensure that the speaker embedding is encoded into cosine space which is more suitable for the scoring and clustering in the following. It works by maximizing the cosine similarity among the embeddings extracted from the same speaker. Finally, the total loss is summarized with coefficient $\alpha$:
\begin{equation}
    L_{dino} = L_{ce} + \alpha \sum_{e\in \{e_1^l,e_2^l\}}\sum_{e'\in \{e_1^l,e_2^l, e_1^s, \dots, e_4^s\}} ( 1 - \frac{e \cdot e'}{\left \| e \right \| \left \| e' \right \| })
\end{equation}
where $e$ represents the extracted speaker embedding from encoder.

\subsection{Cluster-Aware Training on DINO}

For traditional DINO, all segments are sampled from the same utterance to form positive pairs. Limited by the duration of the utterances, these segments usually have a great degree of overlaps. As mentioned above, the optimization of DINO is encourage \textit{short-to-long} correspondences by minimizing the cross-entropy between two distributions of positive pairs. Because there are a lot of overlapped parts in the segments, the model might tend to pay more attention to the content, channel and other irrelevant information of the overlapped parts, and ignore the speaker information in the audio. Although we will add different types of data augmentation to segments, the data still lacks diversity which could lead the model optimization in the wrong direction.

\begin{figure}[ht]
  \centering
  \includegraphics[width=0.85\linewidth]{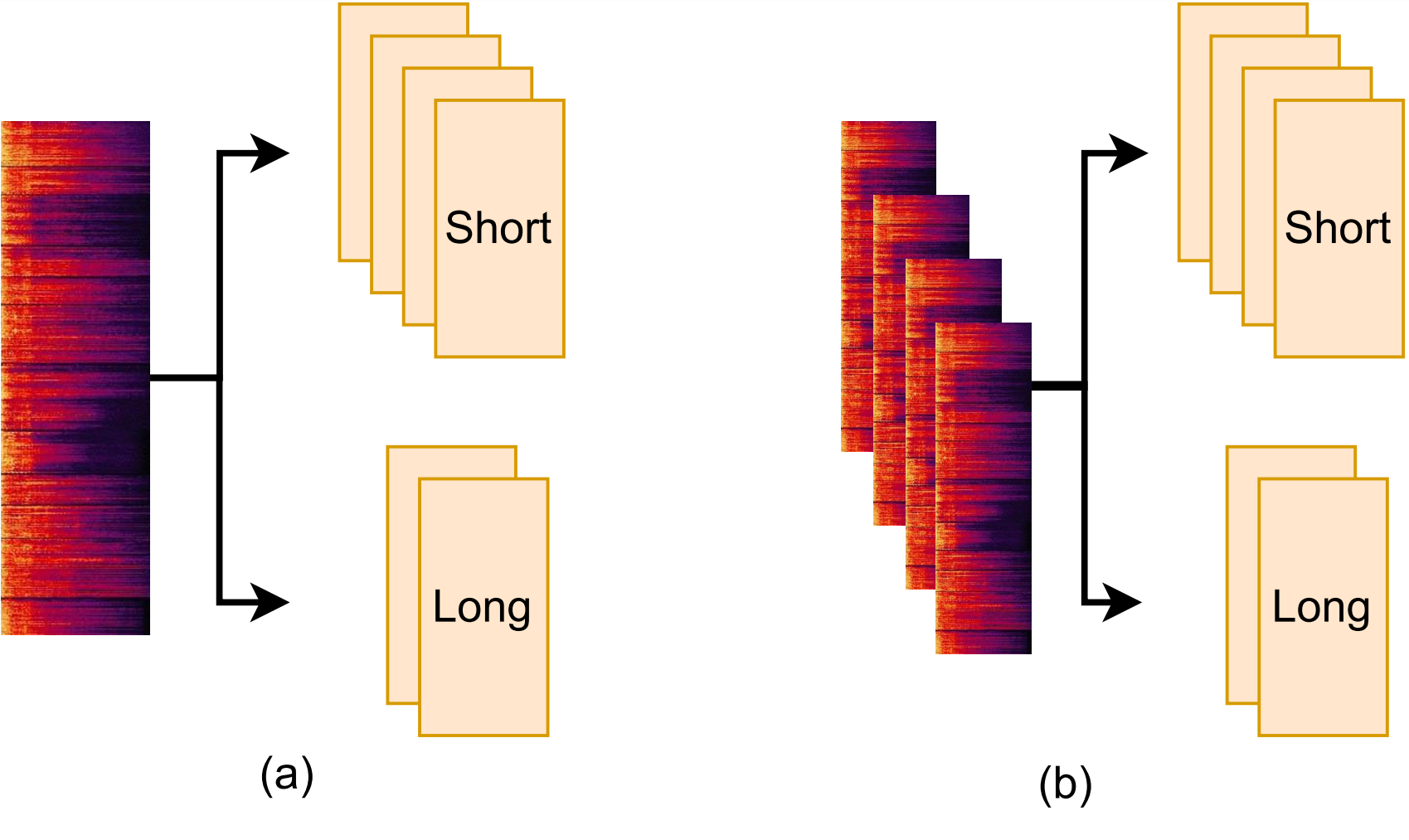}
  \caption{Difference between traditional DINO and cluster-aware training DINO. (a) Traditional DINO:long and short segments are sampled from the same utterance to compose the positive pairs. (b) Cluster-aware training DINO: through simple clustering algorithm, we consider that the same speaker in the same cluster shares the same identity and segments are cropped from the corresponding cluster.}
  \label{fig:cluster_aware}
\end{figure}

In order to reduce the overlaps of segments and increase the diversity of data, we propose a clustering-aware (CA) training strategy for DINO while maintaining the original assumptions as much as possible, which is named CA-DINO in the following. We divide model training into two stages. In the early stage of training, we optimize the model according to the traditional DINO strategy. When the model has the ability to extract discriminative speaker representation, the training process will enter the next stage. In the second stage, the clustering algorithm is performed using the extracted speaker embeddings and we assume that utterances in the same cluster belong to the same person. As shown in Fig.~\ref{fig:cluster_aware}, the positive pairs are sampled from several utterances belonging to the same cluster rather than a single utterance. These pairs come from the same speaker, but with different speaking contents and channels, which leads to a high data diversity and makes the model pay more attention to the speaker's information instead of irrelevant factors. Considering the resource consumption of extracting the embedded speaker, the clustering operation will be done every few rounds.

\section{Iterative Learning with Dynamic Loss-gate and Label Correction}

Based on the proposed CA-DINO self-supervised learning, we then apply the iterative learning framework~\cite{cai2021iterative} to further improve the performance of self-supervised SV. 
During the iterative process, a serious problem is that the generated pseudo labels contain a lot of noises which will confuse and degrade the network. Considering this limitation, several works have been done to select high-quality pseudo labels. In~\cite{cai2021iterative}, an aggressive training method is applied to purify the labels using clustering confidence but achieves minor profit. 
In~\cite{tao2021self}, they conducted a toy experiment and observed that data with lower loss is more reliable. Then, they propose a loss-gate (LG) strategy to select the data with lower loss by setting a fixed threshold and only use these data to update the model. 
With the LG strategy, the system achieved obvious improvement, but the threshold setting in this method is heavily dependent on human experience, and the unreliable data are not fully utilized.

In this section, we will introduce our proposed DLG-LC to adjust the loss-gate threshold dynamically and correct the unreliable pseudo label to fully utilize the data, and then this DLG-LC approach is extended to utilize the multi-modality for further improvements.

\subsection{Dynamic Loss-Gate}


\begin{figure}[t]
  \centering
  \includegraphics[width=0.8\linewidth]{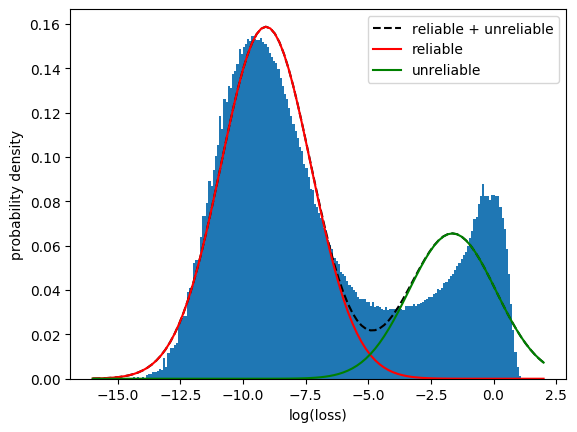}
  \caption{Loss distribution of Loss-gate (LG) learning~\cite{tao2021self} on Voxceleb 2~\cite{voxceleb2}. Loss value is scaled by log function, and the lines are estimated by GMM with two components.}
  \label{fig:loss_distribution}
\end{figure}

In order to determine an appropriate loss-gate threshold, we implemented the LG learning and visualized to analyze the distribution of loss values on Voxceleb 2~\cite{voxceleb2} dataset. 
The histogram of loss values is provided in Fig.~\ref{fig:loss_distribution}. According to the figure, there exist two sharp peaks in the distribution obviously.
And similar experiments conducted in \cite{arazo2019unsupervised} have shown that these data with reliable and unreliable labels can be represented by two peaks respectively.
If we can find a way to model the distribution, then the loss-gate threshold can be determined dynamically as the loss distribution varies, which can avoid laborious manual tuning.

Gaussian distribution is an important continuous probability distribution of real-valued random variables, whose general form of the probability density function is defined in Equation.\ref{equ:gaussian}.
\begin{equation}
\label{equ:gaussian}
\mathcal{N}\left(\mu, \sigma^{2}\right) = \frac{1}{\sigma\sqrt{2\pi}} exp(-\frac{1}{2} (\frac{x-\mu}{\sigma})^2)
\end{equation}
where location parameter and scale parameter are denoted by $\mu$ and $\sigma$ respectively. Gaussian distribution's shape is like a bell, with low on both sides and high in the middle, which is very similar to the ``peaks" of loss in Fig.~\ref{fig:loss_distribution}. In this case, Gaussian Mixture Model (GMM) with two components can be applied to model the loss distribution of reliable and unreliable samples respectively:
\begin{equation}
\label{equ:2_component_gmm}
p(x)=\lambda_1 \mathcal{N}\left(\mu_1, \sigma_1^{2}\right) + \lambda_2 \mathcal{N}\left(\mu_2, \sigma_2^{2}\right)
\end{equation}
where $\lambda_1$ and $\lambda_2$ represent the weights for two Gaussian components.
After fitting, the fitted curves are plotted in Fig.~\ref{fig:loss_distribution}, it's obvious to find that the two weighted Gaussian components can be used to approach these two ``peaks". 
Then, by computing the loss values whose probabilities belonging to the two components are equal, the loss-gate threshold $\tau_1$ can be obtained easily to distinguish between the reliable and unreliable data:
\begin{equation}
\label{equ:obtain_threshold}
\tau_1 : p_1(\tau_1) = p_2(\tau_1)
\end{equation}
where $p_1(x) = \lambda_1 \mathcal{N}\left(\mu_1, \sigma_1^{2}\right)$ and $p_2(x)=\lambda_2 \mathcal{N}\left(\mu_2, \sigma_2^{2}\right)$. For each epoch, all loss values are recorded for re-estimating the parameters of GMM, so $\tau_1$ can be tuned dynamically according to the current training condition.

Our DLG introduces this dynamical loss-gate threshold $\tau_1$ into the speaker classification loss function ArcMargin Softmax (AAM)~\cite{margin_matter} to select the data and only these retained data with losses under the threshold are used to update the parameters of the network.
\begin{equation}
\label{equ:reliable_prob}
    L_{DLG} = \sum_{i=1}^{N} \mathbbm{1}_{l_i < \tau_1}  \log\frac{e^{s(\cos(\theta_{y_i,i}+m))}}{Z}
\end{equation}
where $Z=e^{s(\cos(\theta_{y_i,i}+m))} + \sum_{j=1,j\neq i}^{c} e^{s(\cos(\theta_{y_i,i}))}$, $\theta_{j,i}$ is the angle between the column vector $W_j$ and embedding $x_i$. $s$ is the scaling factor and $m$ is  hyperparameter to control the margin. 
AAM can enforce larger gaps between the nearest speakers and is widely adopted in speaker recognition tasks.

\subsection{Label Correction}


For those unreliable data with large losses, it's wasteful to drop them away directly. Therefore, we propose the label correction (LC) strategy to correct pseudo labels dynamically so that we can utilize the unreliable data effectively.
Researchers in~\cite{tong21_interspeech} have indicated that the network is capable of clustering noisy samples into their correct classes. 
To leverage this ability, we hypothesize that the output prediction of the model is more reliable than pseudo labels generated by clustering. 
Thus the predicted posterior probability is regarded as the target labels and incorporated into the objective loss function to prevent the model from fitting into inaccurate labels.
However, not all prediction labels are suitable for training. Inspired by~\cite{lee2013pseudo, sohn2020fixmatch}, we assume that the prediction label owns high confidence if the model assigns a high probability to one of the possible classes. 
Then, another threshold $\tau_2$ is introduced to retain the prediction whose probability of largest class is above $\tau_2$, and the label correction loss is defined as the following Equation~\ref{equ:lc}:
\begin{equation}
\label{equ:lc}
    L_{LC} = \sum_{i=1}^{N} \mathbbm{1}_{l_i > \tau_1, \max(\hat{p_i})>\tau_2} H(\hat{p_{clean}} \mid p_{aug})
\end{equation}
where $p_{aug}$ represents the output probability of augmented segments and $\hat{p_{clean}}$ represents their corresponding clean version. $H(\cdot)$ here denotes the cross-entropy loss function between two probability distributions. 
In addition, to encourage a peaky distribution, a sharpening operation is applied on $\hat{p_{clean}}$ with sharpness factor $\epsilon_c$ which is described in Equation.~\ref{equ:sharpen}. 

Then, the DLG loss and LC loss are combined to optimize the speaker model as Equation.~\ref{equ:final_loss}.
\begin{equation}
\label{equ:final_loss}
    L = L_{DLG} + L_{LC}
\end{equation}

More specifically, the pseudo-code for describing the flow of the DLG-LC algorithm is provided in detail and shown in Algorithm.~\ref{algo:algorithm}.

\begin{algorithm}[t]
\caption{The proposed Dynamic Loss-Gate and Label Correction}
\label{algo:algorithm}
\LinesNumbered
\footnotesize

\KwIn{mini-batch $D_m=\{(x_1, x_2, y)\}_{i=1}^n$; two threshold $\tau_1$ and $\tau_2$; Network $g(\cdot)$; sharpness factor $\epsilon_c$} 
\KwOut{ the loss of the mini-batch}
\For{$(x_1, x_2, y) \in D_m$}{
    $x_{clean}$, $x_{aug} = $ $x_1$, augment($x_2$) \hspace{5mm} \# augment one segment \\ 
    $p_{clean}$, $p_{aug}$ = $g(x_{clean})$, $g(x_{aug})$ \hspace{2mm} \# output distribution\\ 
    Compute the AAM-softmax loss $l_{clean}$ and $l_{aug}$ according the pseudo label $y$ \\
    Record the $l_{clean}$ value\\ 
    \eIf{$l_{clean}<\tau_1$}{ 
        return $l_{aug}$ \hspace{2cm} \# pseudo label $y$ is reliable \\
    }{
        \eIf{$\max(p_{clean}) > \tau_2$}{
            $\hat{p_{clean}}$ = sharp($p_{clean}$, $\epsilon_c$) \hspace{2mm} \# sharp the distribution\\
            compute the cross-entropy $l$ between $\hat{p_{clean}}$ and $p_{aug}$ \\
            return $l$ \\
        }{
            return 0 \hspace{2cm} \# prediction isn't  reliable\\ 
        }
    }
}
After one epoch, re-estimate the GMM on the recorded loss values and then update the $\tau_1$  \\
\end{algorithm}

\subsection{Incorporate with Multi-Modality}
The researchers in~\cite{cai2022incorporating} have introduced the multi-modality information into the data clustering step to generate more accurate pseudo labels in self-supervised speaker verification. In our work, considering the feature that audio and visual from the same video share the same speaker identity, we also try to add the visual modality to our DLG-LC method for better data utilization, hoping to obtain further improvement. Our fusion of visual information is mainly divided into two aspects: one is to use visual information to help DLG-LC select more reliable data, and the other is to make clustering results better during data clustering.

\subsubsection{multi-modal based DLG-LC}
Different from the single-modal DLG-LC, our strategy of selecting reliable data has been slightly adjusted. For multi-modal data, we will use two independent encoders to encode audio and visual data. Then, through recording the loss values, we can obtain two loss-gate thresholds for audio and visual respectively. For an audio-visual instance, it can be regarded as a reliable label only if its loss values are both under these two loss-gate thresholds, and then we will optimize this instance with AAM softmax which is defined as Equation.~\ref{equ:reliable_prob}.

For unreliable data, the multi-modal label correction will be performed on it. First, we compare whether the predicted labels of the two modal networks are consistent. If the predictions of the two models belong to the same class, it indicates that the accuracy of the prediction is relatively high. Unlike single-mode label correction, which uses soft labels for training, our output is verified by multi-modal, which has higher reliability. As a result, we use the ``hard" labels (i.e. the $\arg\max$ of the model's distribution) as labels to optimize the models by AAM softmax. If the network disagrees with the predicted labels, then we use the soft labels to optimize models respectively.

\subsubsection{multi-modal based data clustering} 
In the previous training step, the multi-modal information was only used to select reliable data, and the models of the two modalities were not structurally related. 
As a result, we can obtain audio $g_a(\cdot)$ and visual encoders $g_v(\cdot)$ independently. 
Given a dataset with audio $x_a$ and visual modality $x_v$, we can use trained encoder to extract audio embedding $e_a$ and visual embedding $e_v$ respectively. 
Considering that the audio and visual embeddings contain complementary information from different modalities, we apply an additional clustering on the joint representation $e_{av} = (e_a, e_v)$, which is formed as the concatenation of audio and visual embeddings. With the joint operation, the representation will be more discriminative and the cluster will be more robust. Then, pseudo labels for the next iteration will be generated by $k$-means on these audio-visual joint embeddings.

\section{Experiments Setup}

\subsection{Dataset}

The experiments are conducted on Voxceleb~\cite{voxceleb1, voxceleb2} which is a large-scale audio-visual dataset for the speaker recognition task. For the model training in stage I and II of self-supervised learning, we adopt the development set of Voxceleb 2~\cite{voxceleb2} for training the networks, and no speaker identity information is used during this process. Because we introduced visual features into the iterative learning stage, we excluded some utterances with the video missing in the data set. Then, the final audio-visual training set comprises $1,091,251$ utterances among $5,994$ speakers, extracted from YouTube.

For the evaluation, we report the experimental results on 3 trials as defined in~\cite{voxceleb2}: the Original, Extended, and Hard Voxceleb test sets. \textbf{Vox-O} is the original test set of Voxceleb 1 contains $37,720$ trials from $40$ speakers.  \textbf{Vox-E} is an trial list which (using the entire dataset) contains $581,480$ trials from $1251$ speakers. \textbf{Vox-H} is a hard evaluation list consisting of $552,536$ pairs sampled from $1190$ speakers in Voxceleb 1, all of which are from the same nationality and gender.

\subsection{Metrics}
The main metrics adopted in this paper are (i) Equal Error Rate (EER) which is the error rate when both acceptance and rejection rates are equal, and (ii) the normalized
minimum Detection Cost Function (minDCF) which is defined by Equation.~\ref{equ:mindcf} : 
\begin{equation}
\label{equ:mindcf}
C_{det} = C_{miss} \times P_{miss} \times P_{tar} + C_{fa} \times P_{fa} \times (1-P_{tar}) 
\end{equation}
where we set the prior target probability $P_{tar}$ as $0.01$ and equal weights between misses $C_{miss}$ and false alarms $C_{fa}$. Both EER and minDCF are commonly used as evaluation metrics for speaker verification systems.

\subsection{Data Augmentation}

\subsubsection{Audio}

To generate extra training samples and increase the diversity of data, we perform online data augmentation strategy~\cite{online_aug} by adding background noise or convolutional reverberation noise from MUSAN~\cite{musan} and RIR dataset~\cite{rir} respectively. The noise types in MUSAN include ambient noise, music, television, and babble noise for the background additive noise. We can obtain augmented data by mixing the noise with the original speech in time-domain waveform directly and the signal-to-noise ratios (SNR) are randomly applied between 5 to 20 dB. For the reverberation, the convolution operation is performed with 40,000 simulated room impulse responses (RIR)~\cite{rir}. After applying the augmentation, we normalize the waveform value for stable training. 
We used 80-dimensional log Mel filter-bank energies with 25ms length Hamming windows and 10ms window shift as the acoustic features, while no voice activity detection (VAD) is involved in our experiments.

\subsubsection{Visual}

For each video segment in VoxCeleb 1 \& 2 datasets, images are extracted at one frame per second. Then, we align the faces in extracted frames using the landmarks predicted by MTCNN ~\cite{mtcnn} and after that, the similarity transformation is used to map the face region to the same shape ($3 \times 112 \times 96$). 
In order to better extract visual features, we convert the image to the most common size of the model ($3 \times 224 \times 224$). And in the following, several data augmentation strategies including random color distortion, random horizontal flipping, random grey scaling, and random Gaussian blur are applied to the original images with a certain probability. 
Finally, we normalize the pixel value of each image to the range of [-0.5, 0.5] before feeding it into the model.

\subsection{CA-DINO Setup}
\subsubsection{DINO} For DINO, considering the training time and memory limitation, we adopt ECAPA-TDNN~\cite{ecapa} as an audio encoder to learn discriminative speaker representation, which is a time-delay neural network (TDNN)~\cite{xvector} based backbone with emphasized channel attention, propagation, and aggregation. It employs a channel- and context- dependent attention mechanism~\cite{gao2019res2net}, Multi-layer Feature Aggregation (MFA), as well as Squeeze-Excitation (SE)~\cite{senet} and residual blocks. The model architecture of ECAPA-TDNN is shown in Table.\ref{tab:ecapa}. 
For each utterance, two long (3 seconds) and four short (2 seconds) segments are randomly cropped and regarded as positive pairs. It is worth noting that all the segments will be applied data augmentation, and after that, they are encoded into 192-dimensional speaker embeddings by the encoder. Similar to the configuration in~\cite{dino}, the $K$ in the DINO projection head is set as $65,536$. Temperatures for the teacher $\epsilon_t$ and the student $\epsilon_s$ are $0.04$ and $0.1$ respectively. In addition, we set cosine loss weight $\alpha$ as $1.0$ to balance two losses. The whole training process will last 150 epochs. Model parameters are updated using stochastic gradient descent (SGD) algorithm with weight decay 5$e$-5. The learning rate is linearly ramped up from 0 to 0.2 in the first 20 epochs, and then it decays to 1$e$-5 with the cosine schedule~\cite{loshchilov2016sgdr}. Moreover, the momentum also follows the cosine schedule from $0.996$ to $1.0$.

\subsubsection{Cluster-Aware Training}
For cluster-aware training strategy, we train the model normally in the first 90 epochs. After that, $k$-means based clustering algorithm is applied on the whole training set every 5 epochs, which is supported by faiss library~\cite{faiss}. The results of clustering are used for the generation of training data. Positive pairs are sampled from utterances belonging to the same cluster rather than the single one.

\begin{table}[ht]
  \caption{Model Architecture of visual encoder ResNet34~\cite{resnet}. $\mathbf{C}$ (kernal size, channel) denotes the convolutional 2D layer. $[\cdot]$ represents the residual block and $L$ is the image size of input. }
  \label{tab:resnet}
  \centering
  \begin{tabular}{lll}
    \toprule
    \textbf{Layer} &  \textbf{Structure}  & \textbf{Output Size} \\
    \midrule
    Input & - & $3\times L \times L$ \\
    Conv2D &  $\mathbf{C}(3 \times 3, 32)$ & $ 32 \times L \times L $ \\
    \midrule
    Residual Block 1 &  $ \begin{bmatrix}
 \mathbf{C}(3\times 3, 32) \\
 \mathbf{C}(3\times 3, 32)
 \end{bmatrix}  \times 3 $, stride 2 & $ 32 \times \frac{L}{2} \times \frac{L}{2}$ \\
    Residual Block 2 &  $ \begin{bmatrix}
 \mathbf{C}(3\times 3, 64) \\
\mathbf{C}(3\times 3, 64)
\end{bmatrix}  \times 4 $ , stride 2 & $64 \times \frac{L}{4} \times \frac{L}{4}$ \\
    Residual Block 3 &  $ \begin{bmatrix}
 \mathbf{C}(3\times 3, 128) \\
\mathbf{C}(3\times 3, 128)
\end{bmatrix}  \times 6 $ , stride 2 & $128 \times \frac{L}{8} \times \frac{L}{8}$ \\
    Residual Block 4 &  $ \begin{bmatrix}
 \mathbf{C}(3\times 3, 256) \\
\mathbf{C}(3\times 3, 256)
\end{bmatrix}  \times 3 $ , stride 2 & $256 \times \frac{L}{16} \times \frac{L}{16}$ \\
    \midrule
    Embedding & - & $192$ \\
    \bottomrule
  \end{tabular}
\end{table}

\subsection{DLG-LC Setup}

\begin{table}[ht]
  \caption{Model Architecture of audio encoder ECAPA-TDNN~\cite{ecapa}. $\mathbf{C}$ (kernal size, channels) denotes the convolutional 1D layer. $F$ is the dimension of the input acoutic features which is determined by the number of frequency bins of the Mel spectrogram. $T$ relates to the frames of the speech segments.}
  \label{tab:ecapa}
  \centering
  \begin{tabular}{lll}
    \toprule
    \textbf{Layer} &  \textbf{Structure}  & \textbf{Output Size} \\
    \midrule
    Input & - & $F \times T$ \\
    Conv1D &  $\mathbf{C}(5, 512)$& $512 \times T$ \\
    \midrule
    SE-Res2Block 1 & \makecell[l]{$\mathbf{C}(1, 512)$ \\ $\mathbf{C}(3, 64) \times 8$,  dilation 2  \\ $\mathbf{C}(1, 512)$} & $512 \times T$ \\
    SE-Res2Block 2 & \makecell[l]{$\mathbf{C}(1, 512)$ \\ $\mathbf{C}(3, 64) \times 8$, dilation 3  \\ $\mathbf{C}(1, 512)$} & $512 \times T$ \\
    SE-Res2Block 3 & \makecell[l]{$\mathbf{C}(1, 512)$ \\ $\mathbf{C}(3, 64) \times 8$, dilation 4  \\ $\mathbf{C}(1, 512)$} & $512 \times T$ \\
    \midrule
    Conv1D & $\mathbf{C}(1, 1536)$ & $ 1536 \times T$ \\
    Pooling Layer & Attentive Stat Pooling & $3072 \times 1$\\
    Embedding & - & $192$ \\
    \bottomrule
  \end{tabular}
\end{table}

\subsubsection{Single Modality}
In this stage, for a fair comparison with ~\cite{tao2021self}, we also adopt ECAPA-TDNN~\cite{ecapa} as our audio encoder to extract speaker embedding. 
For clustering, we choose \textit{k}-means algorithm to assign the pseudo label to the training set. 
Unlike some works~\cite{tao2021self,dku2021,cai2022incorporating} that directly regard the number of real speakers as the number of clusters, we choose $7500$ as the cluster number to verify the robustness of our method.
For label correction, sharpen parameters $\epsilon_c$ and threshold $\tau_2$ are set as $0.1$ and $0.5$ respectively. 
The learning rate decays from $0.1$ to $5e$-$5$ exponentially and we set momentum and weight decay as 0.9 and 1$e$-4. Finally, the training process will last 100 epochs.

\subsubsection{Multi Modality}
For audio-visual based DLG-LC, except for the addition of an image encoder, other configurations are consistent with the single-modal. We employ the ResNet34~\cite{resnet} as the backbone network for the visual encoder, which is similar to the recent works~\cite{groupface,arcface}. More detail is shown in Table.~\ref{tab:resnet}. 

\section{Experimental Results}

The experiments are performed in six parts. 
In section~\ref{sec:exp_ca_dino}, performance comparison of proposed Cluster-aware DINO with previous works in stage I are reported, and we discuss how the number of clusters affects the cluster-aware training strategy.
In section~\ref{sec:exp_finetune}, we report the speaker verification performance of CA-DINO finetuned on the small-scale labeled data. 
In section~\ref{sec:exp_dlg_lc}, an ablation study of our proposed DLG-LC is given to demonstrate its effectiveness.
Then, section~\ref{sec:exp_iteration} and section~\ref{sec:exp_av} show that the proposed dynamic loss-gate and label correction can improve the performance under both single modal and multi-modal scenarios. 
Finally, in section~\ref{sec:exp_final_res}, a comprehensive comparison between our newly proposed self-supervised learning method and previous work demonstrates the superiority and robustness of our system.

\subsection{Evaluation of CA-DINO based Speaker Verification}
\label{sec:exp_ca_dino}
\begin{table}[ht]
  \caption{Performance comparison of the proposed CA-DINO with other self-supervised speaker verification methods. SSL means Self-Supervised Learning. EER (\%) and minDCF (p=0.01) are evaluated on Vox-O test set.}
  \label{tab:dino_result}
  \centering
  \begin{adjustbox}{width=.45\textwidth,center}
  \begin{tabular}{lrr}
    \toprule
    \textbf{SSL Methods} &  \textbf{EER (\%)}  & \textbf{minDCF} \\
    \midrule
    Disent~\cite{nagrani2020disentangled} & 22.090 & - \\
    CDDL~\cite{chung2020seeing} & 17.520 & -\\
    GCL~\cite{inoue2020semi} & 15.260 & -\\
    i-vector~\cite{huh2020augmentation} & 15.280 & 0.63 (p=0.05) \\
    AP + AAT~\cite{huh2020augmentation} & 8.650 & 0.45 (p=0.05) \\
    SimCLR + uniform ~\cite{zhang2021contrastive} & 8.280 & 0.610 \\
    MoCo + WavAug~\cite{xia2021self} & 8.230 & 0.590 \\
    Unif+CEL~\cite{mun2020unsupervised} & 8.010 & - \\
    \midrule
    DINO & 	31.233 & 0.990 \\
    \hspace{1em} + EMA & 4.404 & 0.434 \\
    \hspace{1em} + + Cluster Aware (CA) & \textbf{3.585} & \textbf{0.353} \\
    \bottomrule
  \end{tabular}
  \end{adjustbox}
\end{table}

Table~\ref{tab:dino_result} reports the speaker verification performance of our proposed methods and other previous self-supervised speaker models. 
All the methods are trained on Voxceleb 2 without any speaker label and evaluated on the Vox-O test set. 
According to the results, we can find that the methods based on contrastive learning~\cite{huh2020augmentation,zhang2021contrastive,xia2021self,mun2020unsupervised} have greatly improved the performance compared with the traditional work~\cite{nagrani2020disentangled,chung2020seeing,inoue2020semi}. Our proposed negative-pairs-free CA-DINO achieves a great performance leap again, which shows that negative pairs are indeed a bottleneck for performance improvement. 
In addition, we also provide the ablation study of CA-DINO at the bottom of Table~\ref{tab:dino_result}. 
When we train the DINO without exponential moving average (EMA), it's difficult to converge and only obtains a very bad result which demonstrates that EMA is the key to preventing the model from collapsing. 
And then we apply the cluster-aware (CA) strategy when training the DINO, the performance has been further improved. 
The proposed CA-DINO achieves the EER of \textbf{3.585\%}, with \textbf{55.24\%} relative EER improvement compared with the previously published best performance of self-supervised speaker verification~\cite{mun2020unsupervised}.

\begin{table}[ht]
  \caption{Performance comparison of cluster-aware training with different cluster number. EER (\%) is evaluated on Vox-O test set. 1080k here means that one utterance is one class, which is equivalent to training without cluster-aware strategy.}
  \label{tab:dino_aware_result}
  \centering
  \begin{adjustbox}{width=.45\textwidth,center}
  \begin{tabular}{c|ccccc}
    \toprule
    \textbf{\# Cluster} & 1080k& 30k & 20k & 10k & 5k \\
    \midrule
    \textbf{EER(\%)} & 4.404 & 3.909	& 3.946 & \textbf{3.585} & 3.978 \\
    \bottomrule
  \end{tabular}
  \end{adjustbox}
\end{table}

During the cluster-aware training, there exists a $k$-means clustering operation. 
We also conducted an experiment to explore the influence of the number of clusters on the performance and the results are reported in Table~\ref{tab:dino_aware_result}.
It is observed that our proposed cluster-aware training strategy can bring significant and stable improvements for all the given number of clusters compared with the baseline system (1080k). 
Meanwhile, CA-DINO with 10k cluster number outperforms other systems which shows that the reasonable setting for the number of clusters can maximize the performance improvement.

\subsection{Evaluation of CA-DINO with Pretrain-Finetune Framework with Labeled Data}
\label{sec:exp_finetune}

\begin{table}[ht]
  \caption{EER(\%) comparison of finetuning the pre-trained self-supervised model with different amount of labeled data from Voxceleb 1. Results are evaluated on Vox-O which is the test set of Voxceleb 1.}
  \label{tab:dino_finetune_result}
  \centering
  \begin{adjustbox}{width=.45\textwidth,center}
  \begin{tabular}{lccccc}
    \toprule
    Initial Model & None & 10\% & 20\% & 50\%  & 100\% \\
    \midrule
    Random & 32.78 & 6.893 & 5.276 & 3.691 & 2.755 \\
    SimCLR & 8.547 & 4.388 & 3.797 & 3.266 & 2.936 \\
    CA-DINO & \textbf{3.585} & \textbf{2.393} & \textbf{2.356} & \textbf{2.016} & \textbf{1.835} \\
    \bottomrule
  \end{tabular}
  \end{adjustbox}
\end{table}

In order to better illustrate the superior performance of our proposed CA-DINO, we conduct an exploration of self-supervised learning with pretrain-finetune framework, i.e. fine-tuning the self-supervised model with a small amount of labeled data in the downstream speaker verification task. We randomly sample 10\%/20\%/50\%/100\% labeled utterances from Voxceleb1~\cite{voxceleb1} as the supervision and finetune the self-supervised models with these data. 

From Table \ref{tab:dino_finetune_result}, it is observed that self-supervised model, both SimCLR and proposed CA-DINO, made great improvements compared with model training from scratch, which shows that a pretraining model with better initialization is very important in low-resource conditions. Moreover, comparing the proposed CA-DINO with SimCLR, the proposed non-contrastive CA-DINO outperforms SimCLR obviously and can obtain a good performance position only with few labeled data in downstream speaker verification tasks. Moreover, with only 10\% part of labeled data, CA-DINO even achieves a better performance than the fully supervised system, i.e. 2.393\% vs. 2.755\%, which is meaningful to economize lots of manual annotation.

\subsection{Evaluation of proposed DLG-LC}
\label{sec:exp_dlg_lc}

\begin{table}[ht]
  \caption{EER (\%) comparison on Vox-O, E, H of the proposed DLG-LC in Iteration 1. In this experiment, pseudo labels are estimated from our pre-trained CA-DINO system. SimCLR and CA-DINO here mean we used all the data with the estimated pseudo labels as the supervisory signal without any data selection strategy during the system training.}
  \label{tab:ablation_DLGLC}
  \centering
  \begin{adjustbox}{width=.45\textwidth,center}
  \begin{tabular}{lcccc}
    \toprule
    \textbf{Method} & \textbf{Threshold} & \textbf{Vox-O} & \textbf{Vox-E} & \textbf{Vox-H}  \\
    \midrule
    SimCLR & - & 6.281 & 7.428 & 11.54 \\
    CA-DINO & - & \textbf{2.909} & \textbf{3.315} & \textbf{5.692} \\
    \midrule
    CA-DINO & & & &  \\
    \hspace{1em} + LG~\cite{tao2021self} & 1 & \textbf{2.441} & \textbf{2.930} & \textbf{4.892} \\
    \hspace{1em} + LG~\cite{tao2021self} & 3 & 2.516 & 3.037 & 5.094 \\
    \hspace{1em} + LG~\cite{tao2021self} & 5 & 2.553 & 3.052 & 5.173 \\
    \midrule
    CA-DINO & & & &  \\
    \hspace{1em} + DLG & Dynamic & 2.186 & 2.473 & 4.306 \\
    \hspace{1em} ++ LC & Dynamic & \textbf{2.021} & \textbf{2.331} & \textbf{4.012} \\
    \bottomrule
  \end{tabular}
  \end{adjustbox}
\end{table}

Based on pseudo labels generated by pre-trained models in stage I, we conduct some experiments to illustrate the effectiveness of our proposed methods. The corresponding results are presented in Table~\ref{tab:ablation_DLGLC}.
Firstly, following the iterative learning framework proposed by~\cite{cai2021iterative}, we estimate the pseudo labels based on the speaker embedding extracted by CA-DINO and train a new encoder using these labels. In order to reflect the superiority of our method, we also trained a model based on SimCLR which is the most popular self-supervised speaker verification method~\cite{zhang2021contrastive}. 
From the results in the Table, we can see that the model based on CA-DINO surpasses SimCLR on all test sets with a very large improvement. Then based on pre-trained CA-DINO, we also conduct an exploration of DLG-LC in Iteration 1. 
According to the results, it can be observed that the loss-gate (LG) learning with fixed thresholds to select data can bring significant improvement compared with the system trained without any data selection. It means that loss-gate can effectively select reliable labels which are of benefit to the model. 
However, we also try to set different thresholds (1, 3, 5), and find that the choice of threshold also has a non-negligible impact on model performance~\cite{tao2021self}. 
Based on the estimated GMM, our proposed dynamic loss-gate (DLG) can adjust the threshold dynamically considering the current training situation and obtains better performance than LG which only adopts a fixed threshold during the whole training process. 
In addition, we apply the label correction (LC) strategy to make full use of data with unreliable labels and the results are further improved. Compared with the baseline system (SimCLR without data selection), the proposed CA-DINO with DLG-LC outperforms it by relative \textbf{70.05\%}, \textbf{68.61\%}, \textbf{65.23\%} EER reduction on Vox-O, Vox-E and Vox-H sets respectively.

\subsection{Iterative Learning with DLG-LC}
\label{sec:exp_iteration}

\begin{table*}[ht]
  \caption{EER (\%) and minDCF (p=0.01) comparison on Vox-O, Vox-E, and Vox-H test sets for different iterations of the proposed DLG-LC with other strategies. SimCLR and CA-DINO without DLC-LC mean that we used all the estimated pseudo labels of the data without data selecting in training process.}
  \label{tab:DLGLC_iteration}
  \centering
  \begin{adjustbox}{width=.85\textwidth,center}
  \begin{tabular}{lcccccccc}
    \toprule
    \textbf{Initial Model} & \textbf{DLG-LC} & \textbf{Iteration} & \multicolumn{2}{c}{\textbf{Vox-O}} & \multicolumn{2}{c}{\textbf{Vox-E}} & \multicolumn{2}{c}{\textbf{Vox-H}} \\
    \cmidrule{4-5}
    \cmidrule{6-7}
    \cmidrule{8-9}
    & & & \textbf{EER(\%)} & \textbf{minDCF}  & \textbf{EER(\%)} & \textbf{minDCF} & \textbf{EER(\%)} & \textbf{minDCF}    \\
    \midrule
    \multirow{6}{*}{SimCLR} & \multirow{6}{*}{\ding{53}} & Initial & 8.547 & 0.6453 & 9.228 & 0.6912 & 14.21 & 0.7757 \\
     & & 1 & 6.281 & 0.5811 & 7.428 & 0.6221 & 11.54 & 0.7213 \\
     & & 2 & 5.914 & 0.5299 & 6.745 & 0.5880 & 10.54 & 0.6971 \\
     & & 3 & 5.547 & 0.5259 & 6.407 & 0.5580 & 10.14 & 0.6698 \\
     & & 4 & 4.872 & 0.4651 & 5.593 & 0.5144 & 8.923 & 0.6408 \\
     & & 5 & \textbf{4.484} & \textbf{0.4545} & \textbf{5.225} & \textbf{0.5055} & \textbf{8.501} & \textbf{0.6321} \\
    \midrule
    \multirow{5}{*}{CA-DINO} & \multirow{5}{*}{\ding{53}} & Initial &  3.585 & 0.3529 & 3.852 & 0.4182 & 6.918 & 0.5743 \\
     & & 1 & 2.909 & 0.3000 & 3.315 & 0.3372 & 5.692 & 0.4654 \\
     & & 2 & 2.606 & 0.2887 & 3.181 & 0.3211 & 5.403 & 0.4489 \\
     & & 3 & \textbf{2.558} & 0.3054 & \textbf{3.064} & \textbf{0.3176} & 5.342 & \textbf{0.4482} \\ 
     & & 4 & 2.643 & \textbf{0.2825} & 3.065 & 0.3200 & \textbf{5.291} & 0.4483 \\ 
    \midrule
    \multirow{5}{*}{CA-DINO} & \multirow{5}{*}{\checked} & Initial &  3.585 & 0.3529 & 3.852 & 0.4182 & 6.918 & 0.5743 \\
     & & 1 & 2.021 & 0.2171 &  2.331 & 0.2419 & 4.012 & 0.3484 \\
     & & 2 & 1.596 & 0.1665 & 2.004 & 0.2089 & 3.484 & 0.3083 \\
     & & 3 & \textbf{1.585} & 0.1671 & \textbf{1.879} & \textbf{0.1963} & 3.293 & \textbf{0.2941} \\
     & & 4 & 1.606 & \textbf{0.1636} & 1.906 & 0.2028 & \textbf{3.274} & 0.2955 \\
    \bottomrule
  \end{tabular}
  \end{adjustbox}
\end{table*}

In order to further illustrate the superiority of our proposed method, we carried out several rounds of iterative training following~\cite{cai2021iterative}. 
We summarize the performance of EER and minDCF of each iteration with or without the proposed DLG-LC strategy on Vox-O, Vox-E, Vox-H test sets, and the results are presented in Table~\ref{tab:DLGLC_iteration}. 
Firstly, we compare the iterative results of SimCLR and CA-DINO respectively, and it is noted that both of them are trained without any loss-gate strategies.
According to the results, it is observed that the iterative learning method can continuously improve the performance of the system with the increase of iteration number.
However, the convergence speed based on SimCLR is significantly slower than that based on CA-DINO. SimCLR does not converge even in the 5th round, while CA-DINO has achieved the best performance position in the 3rd round. 
In addition, the final performance of SimCLR with iterative learning is even still worse than the initial performance of CA-DINO. The proposed CA-DINO owns consistent large advantages over SimCLR in each iteration which further demonstrates the superiority of the proposed CA-DINO in self-supervised speaker verification. 

Based on the pseudo-labels generated by CA-DINO, we applied the proposed strategy of DLG-LC, and the performance significantly improved further. 
It only takes one round of iteration to obtain better results than three rounds of iterations without DLC-LC, which shows the importance of the dynamic threshold filtering and label correction on data usage. After convergence with more iterations, its performance is much better than the system without DLG-LC. 
It shows that the proposed DLG-LC can not only speed up the model convergence and reduce the training time but also significantly boost the performance upper limit of the self-supervised learning model.

\subsection{Incorporate with Multi-Modality}
\label{sec:exp_av}

\begin{table*}[ht]
  \caption{EER (\%) and minDCF (p=0.01) comparison on Vox-O, Vox-E, Vox-H test sets for different iterations of the proposed DLG-LC with single- or multi-modality. It's noted that they are both initialed with CA-DINO in the first self-supervised pretraining stage. Both our audio and visual encoders are trained independently, and the fusion of multi-modal information only performs when clustering data and selecting data in iterative learning. We do the testing still with single audio modality.}
  \label{tab:AV_iteration}
  \centering
  \begin{adjustbox}{width=.75\textwidth,center}
  \begin{tabular}{lccccccc}
    \toprule
    \textbf{Training Modality} & \textbf{Iteration} & \multicolumn{2}{c}{\textbf{Vox-O}} & \multicolumn{2}{c}{\textbf{Vox-E}} & \multicolumn{2}{c}{\textbf{Vox-H}} \\
    \cmidrule{3-4}
    \cmidrule{5-6}
    \cmidrule{7-8}
    & & \textbf{EER(\%)} & \textbf{minDCF}  & \textbf{EER(\%)} & \textbf{minDCF} & \textbf{EER(\%)} & \textbf{minDCF}    \\
    \midrule
    Audio & Initial & 3.585 & 0.3529 & 3.852 & 0.4182 & 6.918 & 0.5743 \\
    \midrule
    \multirow{4}{*}{Audio} & 1 & 2.021 & 0.2171 &  2.331 & 0.2419 & 4.012 & 0.3484 \\
     & 2 & 1.596 & 0.1665 & 2.004 & 0.2089 & 3.484 & 0.3083 \\
     & 3 & \textbf{1.585} & 0.1671 & \textbf{1.879} & \textbf{0.1963} & 3.293 & \textbf{0.2941} \\
     & 4 & 1.606 & \textbf{0.1636} & 1.906 & 0.2028 & \textbf{3.274} & 0.2955 \\
     \midrule
    \multirow{3}{*}{Audio-Visual} & 1 & 1.537 & \textbf{0.1326} & 1.789 & 0.1910 & 3.235 & 0.3007 \\ 
     & 2 & \textbf{1.292} & 0.1565 & \textbf{1.571} & \textbf{0.1688} & \textbf{2.799} & \textbf{0.2676} \\
     & 3 & 1.356 & 0.1553 & 1.602 & 0.1711 & 2.839 & 0.2712 \\
    \bottomrule
  \end{tabular}
  \end{adjustbox}
\end{table*}

Then we introduce visual information in the iterative learning process. The difference from the work in~\cite{cai2022incorporating} is that we not only use multi-modality when doing the data clustering but also utilize multi-modality information when applying data selection through DLG-LC. Table~\ref{tab:AV_iteration} illustrates the EER and minDCF performance comparison of DLG-LC with single- and multi-modality. 

It is observed that incorporating both audio-visual modality knowledge in the iterative learning can obtain another large performance, which demonstrates that extra visual information can make the data usage better. Take the EER of Vox-H as an example, with only single modality audio data, the relative EER reduction of the current and previous iterations are \textbf{42.01\%}, \textbf{13.16\%}, and \textbf{5.48\%} on Vox-H trials for the first three iterations. If 
iterative learning with audio-visual data, the relative EER reduction percentages are \textbf{53.24\%}, \textbf{13.48\%} for the first two iterations.

\subsection{Comparison with Other Systems}
\label{sec:exp_final_res}

\begin{table*}[ht]
  \caption{EER (\%) comparison on Vox-O, Vox-E, Vox-H among the proposed CA-DINO with DLG-LC and other most advanced self-supervised systems. The model architecture, clustering number, method and iteration rounds of each system are listed in detail. Noted that AHC and K-M here mean Agglomerative Hierarchical Clustering and $k$-means. ECAPA-S (Small) and ECAPA-L (Large) here denote the ECAPA-TDNN with 512 channels and 1024 channels respectively.}
  \label{tab:DLGLC_overview}
  \centering
  \begin{adjustbox}{width=.99\textwidth,center}
  \begin{tabular}{lccccccc}
    \toprule
    \textbf{Methods} & Model &\textbf{\# Iteration} & \textbf{\# Clusters} & \textbf{Cluster} & \textbf{Vox-O (EER)} & \textbf{Vox-E (EER)} & \textbf{Vox-H (EER)}  \\
    \midrule
    Fully Supervised~\cite{ecapa} & ECAPA-S &- & - & - &  1.010 & 1.240 & 2.320 \\
    \midrule
    IDLab~\cite{idlab2020} & ECAPA-L & 7 & 7500 & AHC & 2.100 & - & - \\
    JHU~\cite{jhu2021} & Res2Net50 & 5 & 7500 & AHC & 1.890 & - & - \\
    SNU~\cite{SNU2021} & ECAPA-L & 5 & 7500 & AHC & 1.660 & - & - \\
    LG~\cite{tao2021self} & ECAPA-L & 5 & 6000 & K-M & 1.660 & 2.180 & 3.760 \\
    DKU + single-modal~\cite{cai2022incorporating} & ResNet34 & 5 & 6000 & K-M & 2.740 & 3.080 & 5.480 \\
    DKU + multi-modal~\cite{cai2022incorporating} & ResNet34 & 5 & 6000 & K-M & 1.920 & 2.030 & 3.720 \\
    \midrule
    CA-DINO & ECAPA-S & 3 & 7500 & K-M & 2.558 & 2.129 & 5.148 \\
    CA-DINO + DLG-LC + single-modal & ECAPA-S & 3 & 7500 & K-M & 1.585 & 1.879 & 3.293 \\
    CA-DINO + DLG-LC + multi-modal & ECAPA-S & 2 & 7500 & K-M & \textbf{1.292} & \textbf{1.571} & \textbf{2.799} \\
    CA-DINO + DLG-LC + multi-modal* & ECAPA-S & 2 & 7500 & K-M & \textbf{1.191} & \textbf{1.474} & \textbf{2.543} \\
    \bottomrule
    \multicolumn{8}{l}{* The results are given with adaptive s-norm~\cite{snorm} for a fair comparison with fully supervised system~\cite{ecapa}.}
  \end{tabular}
  \end{adjustbox}
\end{table*}

In this section, a performance comparison among our proposed CA-DINO with DLG-LC and other self-supervised speaker verification systems is given in Table~\ref{tab:DLGLC_overview}, and most of them are from the latest Voxceleb Speaker Recognition Challenge (VoxSRC)~\cite{nagrani2020voxsrc,brown2022voxsrc} which represent the most advanced systems nowadays. Besides, the fully supervised system is also illustrated as the first line of Table~\ref{tab:DLGLC_overview} for comparison. 


Compared with the previous works using large-size models, the model we adopt is ECAPA-S (Small, C=512) which has fewer parameters and requires fewer computation resources. 
Compared to AHC (Agglomerative Hierarchical Clustering), to make it easier to be implemented, we adopt a simpler and more convenient clustering method K-M ($k$-means) to generate pseudo labels.
Moreover, when clustering data, we set the number of clusters to 7500 instead of 6000, because 6000 is close to the real number of speakers (5994) in the training set which is too special. 
From the results, it's obviously observed that our proposed new self-supervised speaker verification framework is far superior to all the existing methods in both single- and multi-modality, even with fewer iterations, smaller model, and simpler clustering method. For the single modality condition, the proposed CA-DINO with DLG-LC outperforms the best system (LG)~\cite{tao2021self} by relative \textbf{4.52\%}, \textbf{13.81\%} and \textbf{12.42\%} on Vox-O, Vox-E and Vox-H sets respectively with only 3 iterations. If we use audio-visual data in the iterative learning stage, the corresponding improvement is enlarged to relative \textbf{22.17\%}, \textbf{27.94\%} and \textbf{25.56\%}, which is a great performance leap. 

In summary, our proposed system achieves the new \textbf{state-of-the-art} performance for self-supervised speaker verification with a large performance improvement, despite we train the systems with fewer iterations, smaller model, and simpler clustering method. More promisingly, compared to the conventional fully supervised system with ECAPA-TDNN-Small, our newly proposed self-supervised learning system even obtains a comparable performance with the supervised system, but without using any ground-truth labels.

\section{Conclusion}

In this work, we propose an advanced cluster-aware DINO (CA-DINO) with dynamic loss-gate and label correction (DLG-LC) for self-supervised speaker verification. The DINO framework is introduced so that the system can be trained without negative samples, which is greatly improved compared with other self-supervised models. Then cluster-aware training is designed into 
DINO framework, and positive samples are collected from the same category rather than only single sentences, so that the model can utilize more diverse data and obtain system improvement. In the iterative learning stage, a dynamic loss-gate is obtained by modeling the loss histogram with Gaussian distribution, to select reliable data when training on pseudo labels. Instead of dropping unreliable data directly, the predicted posterior is adopted as the target distribution to prevent fitting into incorrect samples. Moreover, multi-modal information is incorporated into DLG-LC to further improve performance. The experiments on Voxceleb show that our newly proposed CA-DINO with DLG-LC is superior and achieves the new \textbf{state-of-the-art} performance for self-supervised speaker verification.
More promisingly, the gap between unsupervised and supervised representation learning is dramatically reduced for speaker verification, and an approaching performance of the fully supervised system is obtained with our self-supervised learning method on speaker verification.


%



\section*{Acknowledgment}
This work was supported in part by China NSFC projects under Grants 62122050 and 62071288, and in part by Shanghai Municipal Science and Technology Major Project under Grant 2021SHZDZX0102. Experiments have been carried out on the PI super-computer at Shanghai Jiao Tong University.

\ifCLASSOPTIONcaptionsoff
  \newpage
\fi



%



\bibliographystyle{IEEEtran}

\bibliography{mybib}

\begin{thebibliography}{10}
\providecommand{\url}[1]{#1}
\csname url@samestyle\endcsname
\providecommand{\newblock}{\relax}
\providecommand{\bibinfo}[2]{#2}
\providecommand{\BIBentrySTDinterwordspacing}{\spaceskip=0pt\relax}
\providecommand{\BIBentryALTinterwordstretchfactor}{4}
\providecommand{\BIBentryALTinterwordspacing}{\spaceskip=\fontdimen2\font plus
\BIBentryALTinterwordstretchfactor\fontdimen3\font minus
  \fontdimen4\font\relax}
\providecommand{\BIBforeignlanguage}[2]{{%
\expandafter\ifx\csname l@#1\endcsname\relax
\typeout{** WARNING: IEEEtran.bst: No hyphenation pattern has been}%
\typeout{** loaded for the language `#1'. Using the pattern for}%
\typeout{** the default language instead.}%
\else
\language=\csname l@#1\endcsname
\fi
#2}}
\providecommand{\BIBdecl}{\relax}
\BIBdecl

\bibitem{han2022self}
B.~Han, Z.~Chen, and Y.~Qian, ``Self-supervised speaker verification using
  dynamic loss-gate and label correction,'' in \emph{Proc. ISCA Interspeech},
  2022.

\bibitem{dvector}
E.~Variani, X.~Lei, E.~McDermott, I.~L. Moreno, and J.~Gonzalez-Dominguez,
  ``Deep neural networks for small footprint text-dependent speaker
  verification,'' in \emph{Proc. IEEE ICASSP}, 2014, pp. 4052--4056.

\bibitem{xvector}
D.~Snyder, D.~Garcia-Romero, G.~Sell, D.~Povey, and S.~Khudanpur, ``X-vectors:
  Robust dnn embeddings for speaker recognition,'' in \emph{Proc. IEEE
  ICASSP}.\hskip 1em plus 0.5em minus 0.4em\relax IEEE, 2018, pp. 5329--5333.

\bibitem{rvector}
H.~Zeinali, S.~Wang, A.~Silnova, P.~Mat{\v{e}}jka, and O.~Plchot, ``But system
  description to voxceleb speaker recognition challenge 2019,'' \emph{arXiv
  preprint arXiv:1910.12592}, 2019.

\bibitem{han2022local}
B.~Han, Z.~Chen, and Y.~Qian, ``Local information modeling with self-attention
  for speaker verification,'' in \emph{Proc. IEEE ICASSP}.\hskip 1em plus 0.5em
  minus 0.4em\relax IEEE, 2022, pp. 6727--6731.

\bibitem{han2022mlp}
B.~Han, Z.~Chen, B.~Liu, and Y.~Qian, ``Mlp-svnet: A multi-layer perceptrons
  based network for speaker verification,'' in \emph{Proc. IEEE ICASSP}.\hskip
  1em plus 0.5em minus 0.4em\relax IEEE, 2022, pp. 7522--7526.

\bibitem{aam}
Y.~Liu, L.~He, and J.~Liu, ``Large margin softmax loss for speaker
  verification,'' \emph{arXiv preprint arXiv:1904.03479}, 2019.

\bibitem{chung2020defence}
J.~S. Chung, J.~Huh, S.~Mun, M.~Lee, H.~S. Heo, S.~Choe, C.~Ham, S.~Jung, B.-J.
  Lee, and I.~Han, ``In defence of metric learning for speaker recognition,''
  \emph{arXiv preprint arXiv:2003.11982}, 2020.

\bibitem{heigold2016end}
G.~Heigold, I.~Moreno, S.~Bengio, and N.~Shazeer, ``End-to-end text-dependent
  speaker verification,'' in \emph{Proc. IEEE ICASSP}.\hskip 1em plus 0.5em
  minus 0.4em\relax IEEE, 2016, pp. 5115--5119.

\bibitem{revising_pooling}
S.~Wang, Y.~Yang, Y.~Qian, and K.~Yu, ``Revisiting the statistics pooling layer
  in deep speaker embedding learning,'' in \emph{2021 12th International
  Symposium on Chinese Spoken Language Processing (ISCSLP)}.\hskip 1em plus
  0.5em minus 0.4em\relax IEEE, 2021, pp. 1--5.

\bibitem{zhu2018self}
Y.~Zhu, T.~Ko, D.~Snyder, B.~Mak, and D.~Povey, ``Self-attentive speaker
  embeddings for text-independent speaker verification.'' in \emph{Proc. ISCA
  Interspeech}, vol. 2018, 2018, pp. 3573--3577.

\bibitem{gmmubm}
D.~A. Reynolds, T.~F. Quatieri, and R.~B. Dunn, ``Speaker verification using
  adapted gaussian mixture models,'' \emph{Digital signal processing}, vol.~10,
  no. 1-3, pp. 19--41, 2000.

\bibitem{ivector}
N.~Dehak, P.~Kenny, R.~Dehak, P.~Dumouchel, and P.~Ouellet, ``Front-end factor
  analysis for speaker verification,'' \emph{IEEE/ACM Trans. ASLP.}, vol.~19,
  no.~4, pp. 788--798, 2011.

\bibitem{baevski2020wav2vec}
A.~Baevski, Y.~Zhou, A.~Mohamed, and M.~Auli, ``wav2vec 2.0: A framework for
  self-supervised learning of speech representations,'' \emph{Proc. NIPS},
  vol.~33, pp. 12\,449--12\,460, 2020.

\bibitem{hsu2021hubert}
W.-N. Hsu, B.~Bolte, Y.-H.~H. Tsai, K.~Lakhotia, R.~Salakhutdinov, and
  A.~Mohamed, ``Hubert: Self-supervised speech representation learning by
  masked prediction of hidden units,'' \emph{IEEE/ACM Trans. ASLP.}, vol.~29,
  pp. 3451--3460, 2021.

\bibitem{fan2020exploring}
Z.~Fan, M.~Li, S.~Zhou, and B.~Xu, ``Exploring wav2vec 2.0 on speaker
  verification and language identification,'' \emph{arXiv preprint
  arXiv:2012.06185}, 2020.

\bibitem{chen2022large}
Z.~Chen, S.~Chen, Y.~Wu, Y.~Qian, C.~Wang, S.~Liu, Y.~Qian, and M.~Zeng,
  ``Large-scale self-supervised speech representation learning for automatic
  speaker verification,'' in \emph{Proc. IEEE ICASSP}.\hskip 1em plus 0.5em
  minus 0.4em\relax IEEE, 2022, pp. 6147--6151.

\bibitem{stafylakis2019self}
T.~Stafylakis, J.~Rohdin, O.~Plchot, P.~Mizera, and L.~Burget,
  ``Self-supervised speaker embeddings,'' \emph{arXiv preprint
  arXiv:1904.03486}, 2019.

\bibitem{inoue2020semi}
N.~Inoue and K.~Goto, ``Semi-supervised contrastive learning with generalized
  contrastive loss and its application to speaker recognition,'' in \emph{Proc.
  IEEE APSIPA ASC}.\hskip 1em plus 0.5em minus 0.4em\relax IEEE, 2020, pp.
  1641--1646.

\bibitem{huh2020augmentation}
J.~Huh, H.~S. Heo, J.~Kang, S.~Watanabe, and J.~S. Chung, ``Augmentation
  adversarial training for unsupervised speaker recognition,'' \emph{arXiv
  preprint arXiv:2007.12085}, 2020.

\bibitem{zhang2021contrastive}
H.~Zhang, Y.~Zou, and H.~Wang, ``Contrastive self-supervised learning for
  text-independent speaker verification,'' in \emph{Proc. IEEE ICASSP}.\hskip
  1em plus 0.5em minus 0.4em\relax IEEE, 2021, pp. 6713--6717.

\bibitem{xia2021self}
W.~Xia, C.~Zhang, C.~Weng, M.~Yu, and D.~Yu, ``Self-supervised text-independent
  speaker verification using prototypical momentum contrastive learning,'' in
  \emph{Proc. IEEE ICASSP}.\hskip 1em plus 0.5em minus 0.4em\relax IEEE, 2021,
  pp. 6723--6727.

\bibitem{mun2020unsupervised}
S.~H. Mun, W.~H. Kang, M.~H. Han, and N.~S. Kim, ``Unsupervised representation
  learning for speaker recognition via contrastive equilibrium learning,''
  \emph{arXiv preprint arXiv:2010.11433}, 2020.

\bibitem{caron2018deep}
M.~Caron, P.~Bojanowski, A.~Joulin, and M.~Douze, ``Deep clustering for
  unsupervised learning of visual features,'' in \emph{Proc. ECCV}, 2018, pp.
  132--149.

\bibitem{cai2021iterative}
D.~Cai, W.~Wang, and M.~Li, ``An iterative framework for self-supervised deep
  speaker representation learning,'' in \emph{Proc. IEEE ICASSP}.\hskip 1em
  plus 0.5em minus 0.4em\relax IEEE, 2021, pp. 6728--6732.

\bibitem{idlab2020}
J.~Thienpondt, B.~Desplanques, and K.~Demuynck, ``The idlab voxceleb speaker
  recognition challenge 2020 system description,'' \emph{arXiv preprint
  arXiv:2010.12468}, 2020.

\bibitem{jhu2021}
J.~Cho, J.~Villalba, and N.~Dehak, ``The jhu submission to voxsrc-21: Track
  3,'' \emph{arXiv preprint arXiv:2109.13425}, 2021.

\bibitem{SNU2021}
S.~H. Mun, M.~H. Han, and N.~S. Kim, ``Snu-hil system for the voxceleb speaker
  recognition challenge 2021,'' \emph{VoxSRC}, 2021.

\bibitem{tao2021self}
R.~Tao, K.~A. Lee, R.~K. Das, V.~Hautam{\"a}ki, and H.~Li, ``Self-supervised
  speaker recognition with loss-gated learning,'' \emph{arXiv preprint
  arXiv:2110.03869}, 2021.

\bibitem{cai2022incorporating}
D.~Cai, W.~Wang, and M.~Li, ``Incorporating visual information in audio based
  self-supervised speaker recognition,'' \emph{IEEE/ACM Trans. ASLP.}, vol.~30,
  pp. 1422--1435, 2022.

\bibitem{dino}
M.~Caron, H.~Touvron, I.~Misra, H.~J{\'e}gou, J.~Mairal, P.~Bojanowski, and
  A.~Joulin, ``Emerging properties in self-supervised vision transformers,'' in
  \emph{Proc. ICCV}, 2021, pp. 9650--9660.

\bibitem{berthelot2019remixmatch}
D.~Berthelot, N.~Carlini, E.~D. Cubuk, A.~Kurakin, K.~Sohn, H.~Zhang, and
  C.~Raffel, ``Remixmatch: Semi-supervised learning with distribution alignment
  and augmentation anchoring,'' \emph{arXiv preprint arXiv:1911.09785}, 2019.

\bibitem{sohn2020fixmatch}
K.~Sohn, D.~Berthelot, N.~Carlini, Z.~Zhang, H.~Zhang, C.~A. Raffel, E.~D.
  Cubuk, A.~Kurakin, and C.-L. Li, ``Fixmatch: Simplifying semi-supervised
  learning with consistency and confidence,'' \emph{Proc. NIPS}, vol.~33, pp.
  596--608, 2020.

\bibitem{zhang2021flexmatch}
B.~Zhang, Y.~Wang, W.~Hou, H.~Wu, J.~Wang, M.~Okumura, and T.~Shinozaki,
  ``Flexmatch: Boosting semi-supervised learning with curriculum pseudo
  labeling,'' \emph{Proc. NIPS}, vol.~34, 2021.

\bibitem{nagrani2020disentangled}
A.~Nagrani, J.~S. Chung, S.~Albanie, and A.~Zisserman, ``Disentangled speech
  embeddings using cross-modal self-supervision,'' in \emph{Proc. IEEE
  ICASSP}.\hskip 1em plus 0.5em minus 0.4em\relax IEEE, 2020, pp. 6829--6833.

\bibitem{chung2020seeing}
S.-W. Chung, H.~G. Kang, and J.~S. Chung, ``Seeing voices and hearing voices:
  learning discriminative embeddings using cross-modal self-supervision,''
  \emph{arXiv preprint arXiv:2004.14326}, 2020.

\bibitem{chen2020simple}
T.~Chen, S.~Kornblith, M.~Norouzi, and G.~Hinton, ``A simple framework for
  contrastive learning of visual representations,'' in \emph{Proc. ICML}.\hskip
  1em plus 0.5em minus 0.4em\relax PMLR, 2020, pp. 1597--1607.

\bibitem{MOCO}
K.~He, H.~Fan, Y.~Wu, S.~Xie, and R.~Girshick, ``Momentum contrast for
  unsupervised visual representation learning,'' in \emph{Proc. CVPR}, June
  2020.

\bibitem{dku2021}
D.~Cai and M.~Li, ``The dku-dukeece system for the self-supervision speaker
  verification task of the 2021 voxceleb speaker recognition challenge,''
  \emph{arXiv preprint arXiv:2109.02853}, 2021.

\bibitem{caron2020unsupervised}
M.~Caron, I.~Misra, J.~Mairal, P.~Goyal, P.~Bojanowski, and A.~Joulin,
  ``Unsupervised learning of visual features by contrasting cluster
  assignments,'' in \emph{Proc. NIPS}, 2020.

\bibitem{hinton2015distilling}
G.~Hinton, O.~Vinyals, J.~Dean \emph{et~al.}, ``Distilling the knowledge in a
  neural network,'' \emph{arXiv preprint arXiv:1503.02531}, vol.~2, no.~7,
  2015.

\bibitem{loshchilov2016sgdr}
I.~Loshchilov and F.~Hutter, ``Sgdr: Stochastic gradient descent with warm
  restarts,'' \emph{arXiv preprint arXiv:1608.03983}, 2016.

\bibitem{voxceleb2}
J.~S. Chung, A.~Nagrani, and A.~Zisserman, ``Voxceleb2: Deep speaker
  recognition,'' in \emph{Proc. ISCA Interspeech}, 2018, pp. 1086--1090.

\bibitem{arazo2019unsupervised}
E.~Arazo, D.~Ortego, P.~Albert, N.~O’Connor, and K.~McGuinness,
  ``Unsupervised label noise modeling and loss correction,'' in \emph{Proc.
  ICML}.\hskip 1em plus 0.5em minus 0.4em\relax PMLR, 2019, pp. 312--321.

\bibitem{margin_matter}
X.~Xiang, S.~Wang, H.~Huang, Y.~Qian, and K.~Yu, ``Margin matters: Towards more
  discriminative deep neural network embeddings for speaker recognition,'' in
  \emph{Proc. IEEE APSIPA ASC}.\hskip 1em plus 0.5em minus 0.4em\relax IEEE,
  2019, pp. 1652--1656.

\bibitem{tong21_interspeech}
F.~Tong, Y.~Liu, S.~Li, J.~Wang, L.~Li, and Q.~Hong, ``{Automatic Error
  Correction for Speaker Embedding Learning with Noisy Labels},'' in
  \emph{Proc. ISCA Interspeech}, 2021, pp. 4628--4632.

\bibitem{lee2013pseudo}
D.-H. Lee \emph{et~al.}, ``Pseudo-label: The simple and efficient
  semi-supervised learning method for deep neural networks,'' in \emph{Proc.
  ICML}, vol.~3, no.~2, 2013, p. 896.

\bibitem{voxceleb1}
A.~Nagrani, J.~S. Chung, and A.~Zisserman, ``Voxceleb: A large-scale speaker
  identification dataset,'' in \emph{Proc. ISCA Interspeech}, 2017, pp.
  2616--2620.

\bibitem{online_aug}
W.~Cai, J.~Chen, J.~Zhang, and M.~Li, ``On-the-fly data loader and
  utterance-level aggregation for speaker and language recognition,''
  \emph{IEEE/ACM Trans. ASLP.}, vol.~28, pp. 1038--1051, 2020.

\bibitem{musan}
D.~Snyder, G.~Chen, and D.~Povey, ``Musan: A music, speech, and noise corpus,''
  \emph{arXiv preprint arXiv:1510.08484}, 2015.

\bibitem{rir}
T.~Ko, V.~Peddinti, D.~Povey, M.~L. Seltzer, and S.~Khudanpur, ``A study on
  data augmentation of reverberant speech for robust speech recognition,'' in
  \emph{Proc. IEEE ICASSP}.\hskip 1em plus 0.5em minus 0.4em\relax IEEE, 2017,
  pp. 5220--5224.

\bibitem{mtcnn}
K.~Zhang, Z.~Zhang, Z.~Li, and Y.~Qiao, ``Joint face detection and alignment
  using multitask cascaded convolutional networks,'' \emph{IEEE signal
  processing letters}, vol.~23, no.~10, pp. 1499--1503, 2016.

\bibitem{ecapa}
B.~Desplanques, J.~Thienpondt, and K.~Demuynck, ``Ecapa-tdnn: Emphasized
  channel attention, propagation and aggregation in tdnn based speaker
  verification,'' \emph{arXiv preprint arXiv:2005.07143}, 2020.

\bibitem{gao2019res2net}
S.~Gao, M.-M. Cheng, K.~Zhao, X.-Y. Zhang, M.-H. Yang, and P.~H. Torr,
  ``Res2net: A new multi-scale backbone architecture,'' \emph{IEEE transactions
  on pattern analysis and machine intelligence}, 2019.

\bibitem{senet}
J.~Hu, L.~Shen, and G.~Sun, ``Squeeze-and-excitation networks,'' in \emph{Proc.
  CVPR}, 2018, pp. 7132--7141.

\bibitem{faiss}
J.~Johnson, M.~Douze, and H.~J{\'e}gou, ``Billion-scale similarity search with
  {GPUs},'' \emph{IEEE Transactions on Big Data}, vol.~7, no.~3, pp. 535--547,
  2019.

\bibitem{resnet}
K.~He, X.~Zhang, S.~Ren, and J.~Sun, ``Deep residual learning for image
  recognition,'' in \emph{Proceedings of the IEEE conference on computer vision
  and pattern recognition}, 2016, pp. 770--778.

\bibitem{groupface}
Y.~Kim, W.~Park, M.-C. Roh, and J.~Shin, ``Groupface: Learning latent groups
  and constructing group-based representations for face recognition,'' in
  \emph{Proc. CVPR}, 2020, pp. 5621--5630.

\bibitem{arcface}
J.~Deng, J.~Guo, N.~Xue, and S.~Zafeiriou, ``Arcface: Additive angular margin
  loss for deep face recognition,'' in \emph{Proc. CVPR}, 2019, pp. 4690--4699.

\bibitem{snorm}
P.~Matějka, O.~Novotný, O.~Plchot, L.~Burget, M.~D. Sánchez, and
  J.~Černocký, ``Analysis of score normalization in multilingual speaker
  recognition,'' in \emph{Proc. ISCA Interspeech}, 2017, pp. 1567--1571.

\bibitem{nagrani2020voxsrc}
A.~Nagrani, J.~S. Chung, J.~Huh, A.~Brown, E.~Coto, W.~Xie, M.~McLaren, D.~A.
  Reynolds, and A.~Zisserman, ``Voxsrc 2020: The second voxceleb speaker
  recognition challenge,'' \emph{arXiv preprint arXiv:2012.06867}, 2020.

\bibitem{brown2022voxsrc}
A.~Brown, J.~Huh, J.~S. Chung, A.~Nagrani, and A.~Zisserman, ``Voxsrc 2021: The
  third voxceleb speaker recognition challenge,'' \emph{arXiv preprint
  arXiv:2201.04583}, 2022.

\end{thebibliography}

\end{document}